\documentstyle[aps,prb,multicol,epsf]{revtex}

\newcommand{\uprule}{\end{multicols}
\noindent \vrule width3.375in height.2pt depth.2pt
\vrule height.5em depth.2pt \hfill \widetext }
\newcommand{\downrule}{\indent \hfill \vrule depth.5em height0pt
\vrule width3.375in height.2pt depth.2pt
\begin{multicols}{2} \narrowtext}

\newcommand{\be}{\begin{equation}}
\newcommand{\ee}{\end{equation}}  
\newcommand{\ba}{\begin{eqnarray}}
\newcommand{\ea}{\end{eqnarray}}

\begin{document}
\draft
\title{Two-component theory of a droplet of electrons in half-filled Landau level}
\author{S.-R. Eric Yang$^{1,2}$ and W.S. Lyue$^{1}$}
\address{Department of Physics, Korea University, Seoul 136-701,
               Korea$^{1}$
                  \\ and \\
Asia Pacific Center for Theoretical Physics, 207-43
              Cheongryangri-dong Dongdaemun-gu Seoul 130-012, Korea$^{2}$}
\maketitle
\begin{abstract}
We have investigated low energy excitations of a disk of electrons in half-filled
Landau level using trail wave function and small-size exact diagonalization approaches.
We have constructed a set of  many-body basis states that describe correctly
the low energy excitations.  In this theory
a droplet consists of {\it two} types of composite fermion liquids, and suggests that
a droplet can support an edge
magnetoplasmon and low energy droplet excitations.
A possibility
of measuring these excitations in a quantum dot is discussed.
\end{abstract}
\begin{multicols}{2}

There has been a great interest in two-dimensional 
electron liquid in a strong magnetic field at half-filling\cite{halp,read1,recent,pas,read2,lee}.
Its bulk properties have been investigated and several novel features have 
been revealed.  Electrons form a new exotic metal in which they move effectively
in zero magnetic field despite the presence of an external field \cite{halp}.  
A dipole-like new particle\cite{read1,recent,read2,lee} with fermionic statistics
has been introduced. 
This new state of matter form a Fermi liquid with a finite compressibility \cite{halpste}.

The edge properties of this exotic liquid are of great current interest, both 
experimentally\cite{gray} and theoretically\cite{shy,Z,yang1,leewen,frad}.
Recently a two-mode model for the edge \cite {leewen} has been proposed 
based on a theory of the electron liquid at $\nu=1/2$ 
consisting of charged bose and neutral Fermi liquids \cite{lee}.
In this theory the charge mode propagate while the neutral mode is dispersion-less.
A theory based topology of Abelian fractional quantum Hall states also leads
to similar results, but in this approach two non-propagating modes of a 
purely topological origin exist\cite{frad}.
The edge properties have been also investigated using a
trial wave function for an  electron disk 
at half-filling\cite{yang1}.
The computed edge occupation number 
showed, in sharp contrast to incompressible Laughlin states, that a tail 
exists in the occupation number beyond the radius of a 
droplet of uniform filling factor $1/2$.  
This result is found to be identical to that of the two-mode theory \cite{yang2}.

A disk of electrons\cite{laug} is an intermediate system between a bulk and an edge.
In many respects it is similar to a nucleus consisting of protons and neutrons.
A  nucleus has several types of collective modes, Mie resonances, geometrical modes,
and surface excitations.
The electron disk at half-filling is different from other quantum Hall liquids 
in that it has a finite compressibility.  It is precisely this that
makes it similar to a nucleus.

In this paper we investigate low energy excitations of a such a disk of electrons. 
In this work we develop a different framework based on
trial wave function approach and not on field theory.  We find that a droplet at half filling
can be viewed as a {\it two-component}  composite fermion (CF) liquid, 
one component consisting of CFs
of higher Landau levels (LLs) and the other of CFs of the 
lowest Landau level (LLL) (see Fig.1(b)).
The main result of our investigation is the identification of low energy  excitations of
the two-component CF liquid. 
One of them is the center of mass (CM) motion, in which the pair excitations 
of each CF liquid move in phase.  This excitation is an edge mode.
We have identified another excitation in which the pair excitation of one CF liquid move out of phase
with the pair excitations of the other CF liquid.  
This is not an edge mode, but is an excitation of the entire droplet.
We find other droplet excitations and discuss the possibility of measuring them
in a quantum dot.

The approach of the present work is based on the following  idea
about the ground state.
In disk geometry at $\nu=1/2$ 
the constraint $\nu=N(N-1)/2M_0$ dictates
that about $N/2$ CFs  occupy the LLL 
and the rest of CFs occupy singly each higher LL \cite{yang1} (Here
$\nu$ is the filling factor and $M_0$ is the total ground state
angular momentum). The energies 
of these singly occupied states
must coincide with the Fermi energy of composite fermions (see Fig.1(b)).
Since collective modes are described by forming a linear combination
of pair excitations in CFLLs, 
these pair excitations
may serve as the basis states.
The number of these basis states at a given wave vector 
is equal to the number of possible pair
excitations. 

We consider a system of 2D electrons confined by a 
parabolic potential , $V(r)=m\Omega^2r^2/2$.    The advantage of this model is
that some exactly known results can be utilized in constructing
the collective modes.
Such a system will form a uniform
density electron disk for sufficiently large $N$.  
We confine our attention here to the 
strong magnetic field limit. 
In this limit the symmetric gauge single-particle
eigenstates are conveniently classified by a Landau level index $n$ and an angular 
momentum index $m\geq n$, and we can confine our attention to $n=0$.
The single particle orbitals in the $n=0$ level 
have energies $\epsilon_m=\hbar\omega_c/2+\gamma(m+1)$,
where $\gamma=m\Omega^2 \ell^2$ ($\ell$ is the magnetic length).   
The system is invariant under spatial rotations about an axis perpendicular to the
2D plane and passing through the center of the dot.  It follows that the total
angular momentum $M_z$ is a good quantum number.  Eigenenergies may be expressed as a sum
of interaction and single-particle contributions
\be
E_i(N,M_z)=U_i(N,M_z)+\gamma(N+M_z).
\ee
Here $i$ labels a state within a $M_z$ subspace, and $U_i(N,M_z)\propto e^2/\epsilon \ell$ is
determined by exactly diagonalizing the electron-electron interaction term in the 
Hamiltonian within this subspace \cite{yang3}.

Recently it has been argued that the main physics of strong electron
correlation is that the electrons carry two Jastrow factors  
\cite{jain}.  
Jain and Kamila\cite{JK} have proposed the following wave functions of interacting
fermions in the LLL for a given total angular momentum $M_{z}$: 
\be
\psi_{LLL}=P_{L^3}Det\{ z_i^{n+m}\partial_{i}^{n}J_i \},
\ee
where $J_i=\prod_{j\not=i}  (z_i-z_j)$.
Successively higher powers of $\partial_i$ acting on $J_i$ appear as we
move to the right for a given row in the determinant\cite{drop}. 
There are $N$ possible values of $(n,m)$ in $\psi_{LLL}$, and the sum of $m$
yields $M_z$.
One may regard $n$ and $m$ as
the LL index and the angular momentum of the CFs, respectively. 

Wave functions of the LLL  
must be analytic\cite{gv}, and the  filling factor is related to the total angular momentum
through $\nu=N(N-1)/2M_z$.  These facts severely limits the possible forms
of the wave functions.  
It has been proposed at
$\nu=1/2$ in disk geometry that a correct ground state wave function $\Phi_0^{'}$ is obtained by
putting about $N/2$ CFs in the LLL and the rest singly in higher LL's\cite{yang1} (In this scheme
CFs are not distributed
like electrons in a Fermi liquid).
For even and odd $N$  states $[N_0,..,N_{i},...,N_{L-1}]$=$[N/2,2,1,...,1]$  
and $[(N+1)/2,1,...,1]$ are chosen, respectively,
where $N_i$ is the number of CFs occupying
the i-th LL.
For even $N$ the number of composite fermions occupying LLL and
higher LL's are, respectively, $N_0=N/2$ and $N_h=N/2$.  
For odd  $N$ they are $(N+1)/2$ and $(N-1)/2$.  

To investigate excited states we make the following assumption: The states
$\Psi_i=[N_0,..,N_{i}^{\ast},...,N_{L-1}]$, where the highest angular momentum
of the $n=i$ LL has increased by unity,
form a basis for the collective modes with $\delta M_z=1$.  The number of states in this basis 
is $N_L$, equal to the number of LL's occupied
by composite fermions (It is given by $N/2$ and $(N+1)/2$ for even and
odd number of $N$). There are many other possible states
that could be added to the basis, but we assume that these states are enough.
These are non-orthogonal states.  Below we test these ideas against 
exact diagonalization results. 

The first collective mode we construct from these states is 
the center of mass motion of
electrons. 
The center-of-mass (CM) motion of a droplet  with $\delta M_z=1$ is described by multiplying the
ground state wave function with powers of the CM coordinate\cite{trugman}
$Z=(\sum_{i=1}^{N} z_i)/N$: 
$\Phi_1^{'}=Z\Phi_0^{'}$. In the absence of the external  
potential, the
displaced state is degenerate in energy with the original state because 
the interaction Hamiltonian, projected to the LLL,
commutes\cite{trugman} with $Z$.
It can be
explicitly confirmed for $\nu=2/5\,\, (N=4)$, $\nu=1/2\,\, (N=3,4,5)$ and
$\nu=2/3\,\, (N=4,5)$ that the following relation holds\cite{han}
\begin{equation}
\Phi_1^{'}=\sum_{i=0}^{N_L-1} \Psi_i.
\label{eq:CMmode}
\end{equation}  
$\Phi_1^{'}$ is represented remarkably simply in our basis:
it is expressed as a linear combination of the
intra-Landau-level excitations of CFs with the same expansion coefficients. 

In order to find another eigenstate  with $\delta M_z=1$ we try a wave function of the form
\be
\Phi_2^{'}=\Psi_1-a\sum_{i=1}^{N_L-1} \Psi_i=\Psi_I- \Psi_{II}.
\label{eq:Cmode}
\ee
The orthogonality of $\Phi_1^{'}$ with $\Phi_2^{'}$ determines the parameter $a$
\be
a=\frac{\sum_{i=0}^{N_L}<\Psi_i|\Psi_1>}{\sum_{i=0}^{N_L-1}\sum_{j=1}^{N_L-1}<\Psi_i|\Psi_j>}.
\ee
For $N=3,4,5$ the parameter $a=2,1,3/2$, which is precisely the ratio between the number of
composite fermions in the lowest LL ($N_0$) and in higher LLs ($N_h$).  
In the limit $N\rightarrow \infty$
we expect $a=N_0/N_h \rightarrow 1$. 

Other $N_L-2$ eigenstates  may be constructed using the Gram-Schmidt process.
The $J$'th unnormalized eigenstate is given by 
\be
\Phi_J^{'}=\Psi_J-<\Phi_1|\Psi_J>\Phi_1-...-<\Phi_{J-1}|\Psi_J>\Phi_{J-1}
\ee
Then $\Phi_J$ is the unit vector $\Phi_J^{'}/<\Phi_J^{'}|\Phi_J^{'}>$.

These wave functions are proposed at arbitrary values of $N$.
They can be  compared with the exact eigenstates
for $N=3,4,5$, and we find a good agreement.  
For $N=5$ there are three $\Phi_J$ states $\Phi_1^{'}=\Psi_1+\Psi_2+\Psi_3 $,
$\Phi_2^{'}=\Psi_1-3/2(\Psi_2+\Psi_3) $, and $\Phi_3^{'}=-0.0654\Psi_1-0.8037\Psi_2+\Psi_3 $.
We find that the overlaps of the normalized $\Phi_1,\Phi_2,\Phi_3$ with the exact wave functions
are all nearly one.
This comparison is done by expanding 
$\Phi_J$ into Slater determinant basis states of the form 
$\Phi_J=\sum_i b_i\phi_i$,
where
$\phi_i=\prod_p a^+_{m_p}|0>$
with $\sum_{p=1}^N m_p=M_z$.  The states $\Phi_J$ can be written as linear combinations
of 34 Slater determinant states.
In $\Phi_1$ particle-hole excitations of the LLL and those of higher
LLLs move in phase.  This implies that composite fermion liquids of higher
LLs and the LLL move in phase.  In  $\Phi_2$ the particle-hole excitation 
of lowest CFLL has opposite
phase in comparison with those of higher CFLLs, which means 
composite fermion liquids of higher
LLs and the LLL move out of phase.
In $\Phi_3$ the amplitude of
the particle-hole excitation of the lowest CFLL is rather small, indicating
that it mainly consists of excitations in the composite fermion liquid of higher
LLs. 

The states $\Phi_1$, $\Phi_2$, and $\Phi_3$  
are responsible for the lowest, second lowest, and third lowest energy peaks
in the plot of the dynamic structure factor for $N=5$  (see Fig.2 ).
Figure 3  displays the occupation numbers $n_m=<c_m^+c_m>$ of these excitations.
In the CM mode only the occupation numbers near the edge are 
changed, indicating an edge mode.
In this mode the electrons simply move back and forth uniformly with
a circular volume, resulting in an edge magnetoplasmon. The energy required to
excite this mode from the ground state 
is given by the intra LL energy spacing $\gamma$, i.e. the excitation
energy is independent of electron-electron interaction. 
In the other excitations 
$n_m$ changes significantly also in the interior of the droplet.

Figure 4 displays the dynamic structure factor for $N=8$.  The biggest
peak represents the CM mode.  Around this main peak several smaller peaks 
separated from the continuum are observed.  These peaks originate from states $\Phi_J$ and 
have much bigger oscillator strength than continuum states.
The excitation
energies of these peaks will contain contributions from electron-electron interaction.
Our small-size exact diagonalization shows that the oscillator strength of these
peaks depend sensitively on $N$.  In the large $N$ limit these $\Phi_J$
excitations are expected to form a continuum and the CM mode will be damped by it, much
like what Halperin et al\cite{halp} find.

Far-infrared radiation can induce optical transition 
between meV energy levels in a quantum dot\cite{Heit}, and is a possible tool
to investigate these excitations.
It will be interesting to observe how these peaks evolve
as $N$ is varied from a mesoscopic number to a macroscopic number.

This work has been supported by
the Ministry of Education
under Grant No. BSRI-96-2444 and by the KOSEF under grant 981-0207-026-2.

\end{multicols}

\begin{figure}
\caption{(a) It displays the occupation of composite fermions in the ground state
for $N=5$.  According to the proposed  trail wave function\cite{yang1}
about half of the composite fermions occupy the lowest Landau level, while the other half
occupy singly higher Landau levels.   
(b)A schematic energy diagram of composite fermion
Landau levels as a function of the distance from the center of the droplet.  
This result is obtained from the following
reasoning.  The composite fermions that occupy higher LLs singly have the quantum
numbers $(n,m)=(n,-n)$.  So the composite fermion in the nth LL is strongly
localized to within $\propto l$ of a ring with radius
$r_n=\ell\sqrt{2(n+1)}$.  As $n$ increases the radius increases.
The size of the droplet of composite fermions in higher LLs
is thus about same as that of the droplet in the LLL.  Also 
the energies of the occupied higher Landau level states
must be degenerate since non-equilibrium population is not allowed.
Three pair excitations are shown.
(c) Three basis states for $N=5$ are shown.  They belong to the 
many-body subspace with the total
angular momentum $M_z=M_0+1$, where $M_0=N(N-1)$ is the ground 
state angular momentum.
Note that each basis state has a particle-hole excitation in a composite
fermion Landau level.  In lowest Landau level, although 
many composite fermions are present, 
only one pair excitation is possible with the correct angular momentum.}
\end{figure}

\begin{figure}
\caption{Dynamic structure factor at $\delta M_z=1$
as a function of the interaction energy
$U_i$.  The actual excitation energy is $E_{i0}=U_i+\gamma-U_0$
(energy is measured in units of $e^2/\epsilon \ell$). 
The number of electrons is $N=5$.}
\end{figure}

\begin{figure}
\caption{Occupation numbers for the ground state (circle), $\Phi_1$ edge magnetoplasmon (square),
$\Phi_2$ droplet excitation (diamond), and $\Phi_3$  droplet excitation (triangle).}
\end{figure}

\begin{figure}
\caption{Same as in Fig.2 with $N=8$.}
\end{figure}


\begin{references} 
\bibitem{halp}B.I. Halperin, P.A.Lee, and N. Read, Phys. Rev.  {\bf47}, 7312 (1993)
\bibitem{read1} N. Read, Semicond. Sci. Technol. {\bf 9}, 1859 (1994);
E. Rezayi and N. Read, Phys. Rev. Lett. {\bf 72}, 900 (1994). 
\bibitem{recent} R. Shankar and Ganpathy Murthy, Phys. Rev. Lett. {\bf
79}, 4437 (1997).
\bibitem{pas}V. Pasquier and F. D. M.
Haldane, Nucl. Phys. B {\bf516}, 719 (1998).
\bibitem{read2}N. Read, cond-mat/9804294.
\bibitem{lee} D. H. Lee, Phys. Rev. Lett. {\bf 80} 4745 (1998).
\bibitem{halpste}B. I. Halperin and Ady Stern, Phys. Rev. Lett. {\bf
80}, 5457 (1998).
\bibitem{gray}M. Grayson, D,C, Tsui, L.N. Pfeiffer, K.W. West, and A.M. Chang,
Phys. Rev. Lett. {\bf 80} 1062 (1998).
\bibitem{shy}A.V. Shytov, L.S. Levitov, and B.I. Halperin, 
Phys. Rev. Lett. {\bf 80}, 141 (1998).
\bibitem{Z}Z\"{u}licke and A. H. MacDonald, cond-mat/9802019.
\bibitem{yang1} S.-R. Eric Yang and J. H. Han, Phys. Rev. B {\bf 57},
R12681 (1998).
\bibitem{leewen}D. H. Lee and X.G. Wen; cond-mat/9809160.
\bibitem{frad}A. Lopez and E. Fradkin, cond-mat/9810168.
\bibitem{yang2}J. H. Han and S.-R. Eric Yang , cond-mat/9810267.
\bibitem{laug}R. B. Laughlin, Phys. Rev. Lett. {\bf
50}, 1395 (1983).
\bibitem{yang3}S.-R. Eric Yang, A.H. MacDonald, M. D. Johnson, Phys. Rev. Lett. {\bf 71}, 3194 (1993).
\bibitem{jain}J. K. Jain, Phys. Rev. Lett. {\bf 63},  199 (1989).
\bibitem{JK} J. K. Jain and R. K. Kamilla, Int. J. Mod. Phys. B {\bf 11}, 2621 (1997).
\bibitem{gv} S. M. Girvin and T. Jach, Phys. Rev. B {\bf 29}, 5617 (1984).
\bibitem{drop} We will drop the universal factor 
$\prod_i e^{-|z_i |^2 /4}$ in this paper. 
\bibitem{trugman} S. A. Trugman and S. Kivelson, Phys. Rev. B {\bf 31},
5280 (1985).
\bibitem{han}J. H. Han and S.R. Eric Yang, Phys. Rev. B {\bf 58}, 
R10163 (1998).
\bibitem{Heit}For a review see D. Heitman and J. P. Kotthaus, Physics Today, June, p.56 (1993).

\end{references}
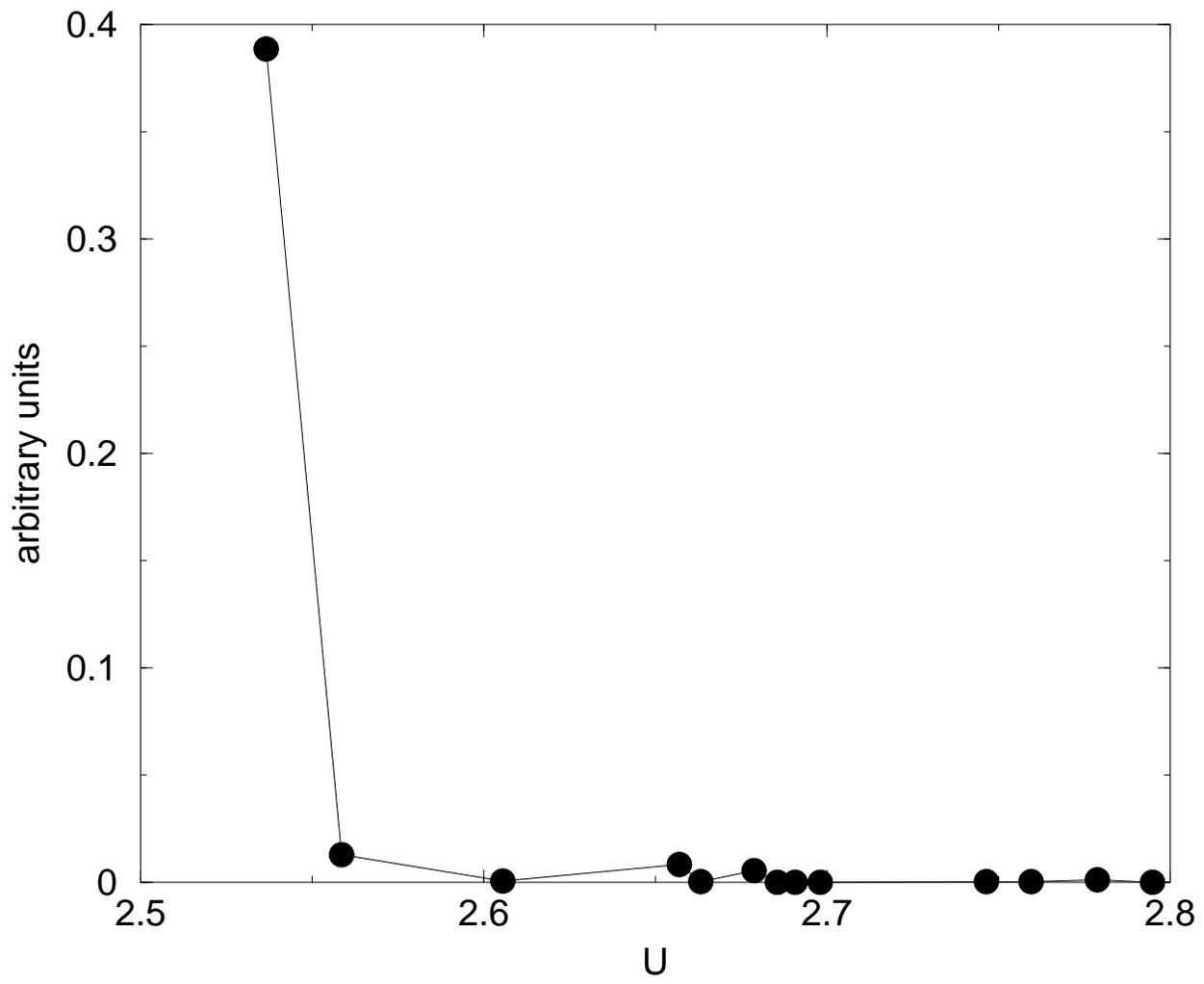
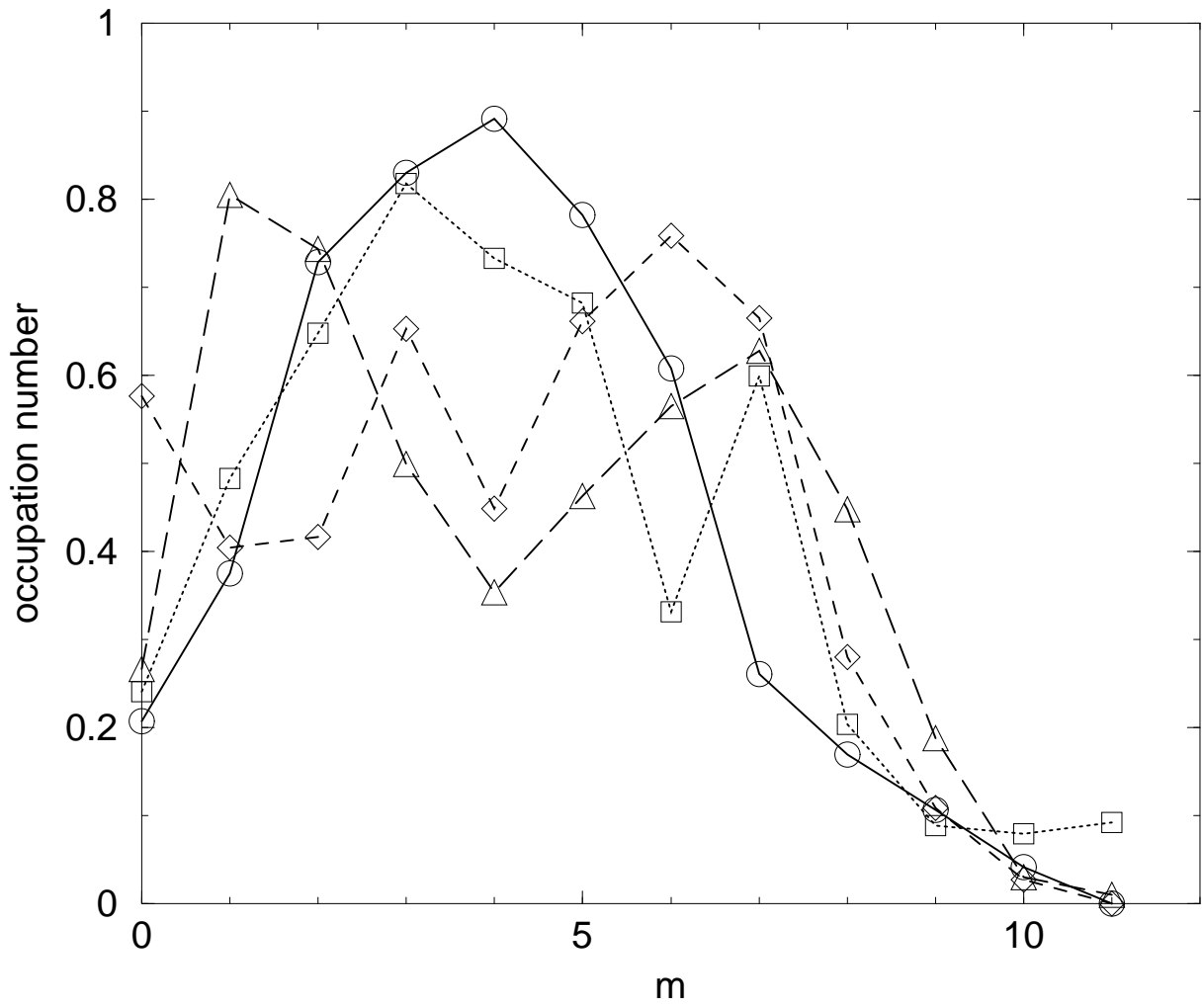
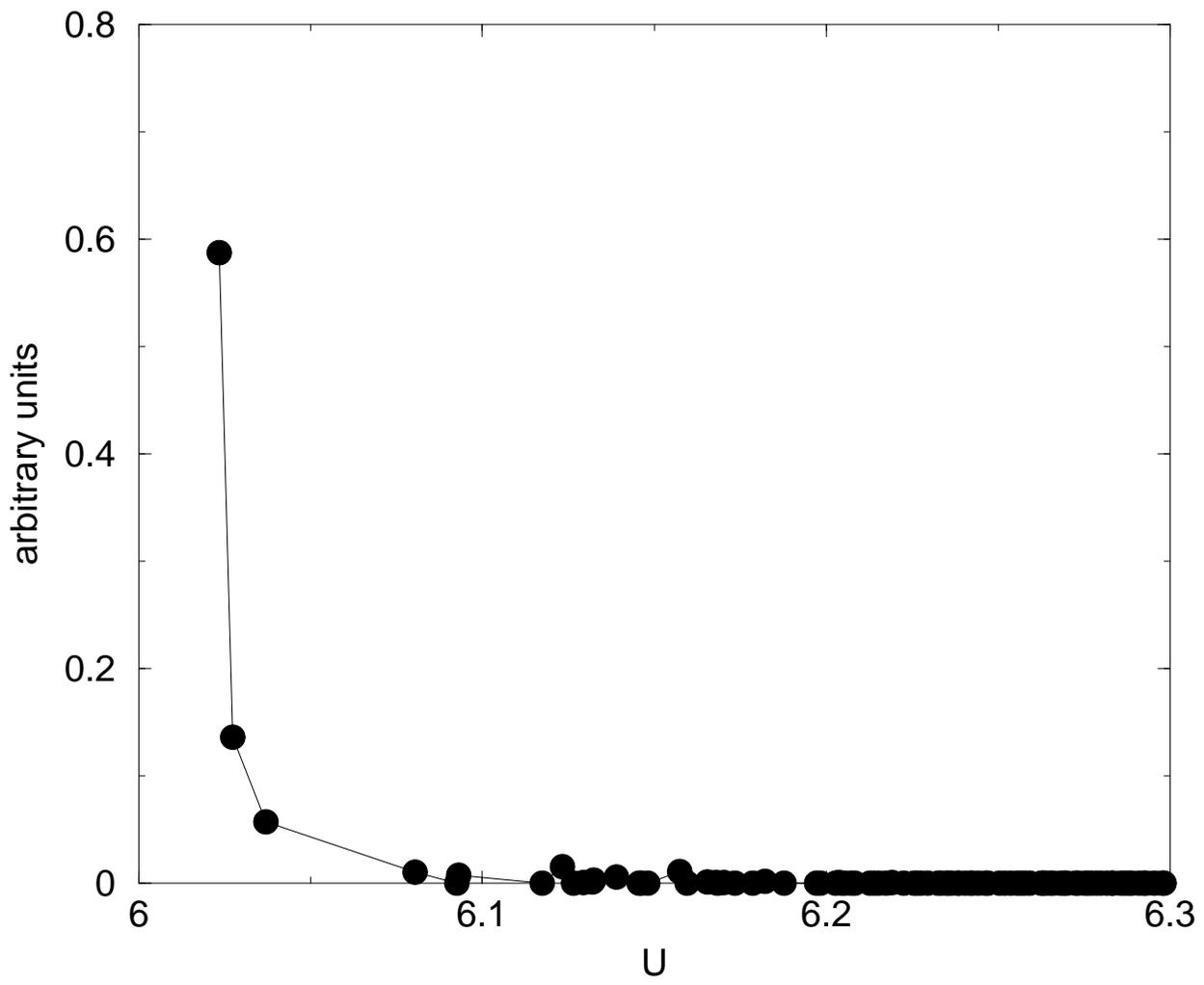
\end{document}